\documentclass[12pt]{article}
\usepackage{epsf}
\usepackage{amsmath,amssymb}

\usepackage{graphicx}

\usepackage{comment}

\begin{document}

\newcommand{\lsim}{\stackrel{<}{_\sim}}
\newcommand{\gsim}{\stackrel{>}{_\sim}}

\newcommand{\rem}[1]{{$\spadesuit$\bf #1$\spadesuit$}}

\renewcommand{\thefootnote}{\fnsymbol{footnote}}
\setcounter{footnote}{0}

\begin{titlepage}

\def\thefootnote{\fnsymbol{footnote}}

\begin{center}

\hfill UT-15-34, IPMU 15-0160, KEK-TH-1863, KEK-Cosmo-183\\
\hfill September, 2015\\

\vskip .75in

{\large \bf 

  Revisiting Big-Bang Nucleosynthesis Constraints \\
  on Dark-Matter Annihilation

}

\vskip .75in

{\large 
  Masahiro Kawasaki$^{(a,b)}$, Kazunori Kohri$^{(c,d)}$, 
  Takeo Moroi$^{(b,e)}$,\\ and Yoshitaro Takaesu$^{(e)}$
}

\vskip 0.25in

\vskip 0.25in

$^{(a)}$
{\em Institute for Cosmic Ray Research, The University of Tokyo, Kashiwa
277-8582, Japan}

\vspace{2mm}

$^{(b)}$
{\em Kavli IPMU (WPI), UTIAS, The University of Tokyo, Kashiwa 277-8583, Japan}

\vspace{2mm}

$^{(c)}$
{\em Theory Center, IPNS, KEK, Tsukuba 305-0801, Japan}

\vspace{2mm}

$^{(d)}$
{\em Sokendai, Tsukuba 305-0801, Japan}

\vspace{2mm}

$^{(e)}$
{\em Department of Physics, University of Tokyo,
Tokyo 113-0033, Japan}

\end{center}
\vskip .5in

\begin{abstract}

  We study the effects of dark-matter annihilation during the epoch of
  big-bang nucleosynthesis on the primordial abundances of light
  elements.  We improve the calculation of the light-element
  abundances by taking into account the effects of anti-nucleons
  emitted by the annihilation of dark matter and the interconversion
  reactions of neutron and proton at inelastic scatterings of
  energetic nucleons.  Comparing the theoretical prediction of the
  primordial light-element abundances with the latest observational
  constraints, we derive upper bounds on the dark-matter
  pair-annihilation cross section.  Implication to some of
  particle-physics models are also discussed.

\end{abstract}

\end{titlepage}

\renewcommand{\thepage}{\arabic{page}}
\setcounter{page}{1}
\renewcommand{\thefootnote}{\#\arabic{footnote}}
\setcounter{footnote}{0}

Recent cosmological and astrophysical observations have revealed
that about $26\%$ of the mass density of the present
universe is occupied by dark matter~\cite{Ade:2015xua}.  This fact
suggests that there exists a stable (or very long-lived) particle or
field which behaves as a non-relativistic object in the present
universe.  However, the particle-physics nature of dark matter is
almost unknown yet, and it is an important task to acquire information
about it.

From particle-physics point of view, a stable particle has attracted
many attentions as a candidate for dark matter; in our analysis, we
assume that some stable particle plays the role of dark matter and
denote it as $X$.  In order for an efficient production of the
dark-matter particle in the early universe, it is often the case that
$X$ has interactions with some of the standard-model particles.  In such a
case, we may derive constraints on the properties of dark matter by
studying the effects of dark-matter pair annihilation into
standard-model particles in the early universe.  

One of the important effects of dark-matter pair-annihilation is on
the big-bang nucleosynthesis (BBN).  If dark matter pair-annihilates
into charged particles during or after the BBN, they induce
photodissociation processes of light elements synthesized via the BBN
reactions.  In addition, energetic hadrons are often produced as a
consequence of the pair-annihilation; if so, hadrodissociation of the
light elements are induced.  Because the standard BBN scenario
predicts the light-element abundances which are more-or-less
consistent with observations, too large pair annihilation cross
section is excluded.  Indeed, such a constraint has been intensively
discussed in literatures \cite{Reno:1987qw, Frieman:1989fx,
  Jedamzik:2004ip, Hisano:2008ti,Hisano:2009rc, Hisano:2011dc,
  Henning:2012rm}.\footnote{%
Besides photodissociation and hadrodissociation,
annihilation of light dark matter ($m_X \lesssim O(10)$~MeV) affects 
the BBN by changing the temperature ratio between neutrinos and 
photons, from which constraints are obtained~\cite{Nollett:2013pwa}.
}

Recently, there have been progresses in the observational determination
of the primordial abundances of the light elements.  In particular,
uncertainty in the primordial deuterium (D) abundance has been
significantly reduced.  Such a progress has a large impact on the BBN
bounds on the dark-matter properties.

In this letter, we revisit the BBN constraint on the annihilation
cross section of dark matter, taking into account the recent
progresses in the observation of the light-element abundances.  For a
reliable calculation of the light-element abundances, we improve the
treatment of hadrodissociation; in particular, we have newly included
the effects of anti-nucleons emitted by the annihilation of dark
matter as well as the effects of interconversion between neutron and
proton at inelastic scatterings of energetic nucleons.  Then,
comparing the theoretical predictions with observational constraints
on the light-element abundances, we derive upper bounds on the
annihilation cross section of dark matter.

We first summarize the current status of the observational constraints
on the primordial abundances of light elements.  The primordial
abundance of D is inferred from D absorption in damped
Ly$\alpha$ systems (DLAs).  Recently Cook {\it et
  al.}~\cite{Cooke:2013cba} observed a DLA toward QSO SDSS J1358+6522
and performed very precise measurement of D.  They also reanalyzed
other four previously known DLAs and, using the total five DLA
samples, obtained the primordial D abundance as
\begin{equation} 
  ({\rm D}/{\rm H})_p = (2.53 \pm 0.04) \times 10^{-5},
\end{equation}
where (A/B) denotes the ratio of number densities of light elements A
and B, and $p$ indicates the primordial value.  Notice that the error
is smaller by a factor of $5$ than that adopted in our previous
study~\cite{Hisano:2009rc}.  This progress in the measurement of D
leads to more stringent constraints on dark-matter annihilation as
seen later.

As for the primordial mass fraction, $Y_p$, of helium~4 ($^4$He), a
new determination with the use of the infrared as well as visible
$^4$He emission lines in 45 extragalactic HII regions was reported in
Ref.~\cite{Izotov:2014fga}, where $Y_p = 0.2551 \pm 0.0022$ is
obtained.  More recently, Aver, Olive and Skillman~\cite{Aver:2015iza}
reanalyzed the data of Ref.~\cite{Izotov:2014fga} and estimated the
$^4$He abundance using Markov chain Monte-Carlo analysis.  They
obtained
\begin{equation}
  Y_p = 0.2449 \pm 0.0040,
\end{equation}
which we adopt in this letter.

We also use a constraint on $^3$He/D which is derived from D and
$^3$He abundances observed in protosolar clouds~\cite{GG03}; taking
into account that the ratio $^3$He/D increases monotonically in time,
we adopt
\begin{equation}
  ({\rm ^3He}/{\rm D})_p < 0.83 + 0.27.
\end{equation}
This observational constraint is the same adopted in
Ref.~\cite{Hisano:2009rc}.

In the previous studies, constraint based on the lithium 7 ($^7$Li)
abundance was also discussed.  However, the situation of the $^7$Li
observation is now confusing.  The observed $^7$Li abundances in
metal-poor halo stars showed almost a constant value
($\log_{10}(^7{\rm Li}/{\rm H}) \simeq -9.8$) called Spite plateau
which was considered as primordial.  However, the recent observation
found much smaller $^7$Li abundances ($\log_{10}(^7{\rm Li}/{\rm H}) <
-10$) for more metal-poor stars~\cite{Sbordone:2010zi}.  Since we do
not know any mechanism to explain such small abundances, we do not use
$^7$Li to constrain the properties of dark matter in this letter.  We
do not use $^6$Li either because $^6$Li abundance is observed as the
ratio to the number density of $^7$Li.

In order to derive constraints on the dark-matter properties, we
calculate the primordial abundances of the light elements, taking into
account the effects of dark-matter annihilation.  Our calculation of
the light-element abundances is based on
Refs.~\cite{KKM,Kawasaki:2008qe} with the modifications explained
below.  The Boltzmann equations for the evolution of the light-element
abundances have the following form:
\begin{align}
  \frac{d n_{A}}{dt} + 3 H n_A = 
  \left[ \frac{d n_{A}}{dt} \right]_{\rm SBBN}
  + \left[ \frac{d n_{A}}{dt} \right]_{\rm photodis}
  + \left[ \frac{d n_{A}}{dt} \right]_{\rm hadrodis}
  + \left[ \frac{d n_{A}}{dt} \right]_{p\leftrightarrow n},
\end{align}
where $n_{A}$ denotes the number density of the light element $A$, and
$H$ is the expansion rate of the universe.  Here, $[dn_A/dt]_{\rm
  SBBN}$ denotes the effects of the standard BBN reactions while the other
terms in the right-hand side are the effects of dark-matter annihilation,
i.e.,  those of photodissociation, hadrodissociation, and
$p\leftrightarrow n$ conversion.  The reaction rates due to the
dark-matter annihilation are proportional to the annihilation rate of
dark matter which is given by\footnote
{ In a series of papers~\cite{KKM,Kawasaki:2008qe}, the effects of
  decaying particles had been studied.  The Boltzmann equations used
  in the present analysis can be obtained from those in
  Ref.~\cite{KKM,Kawasaki:2008qe} by replacing $\Gamma_{\rm
    decay}\rightarrow \Gamma_{\rm annihilation}$, where $\Gamma_{\rm
    decay}$ is the decay rate defined in
  Refs.~\cite{KKM,Kawasaki:2008qe}, with properly rescaling the
  numbers of final-state particles produced by the pair annihilation
  of dark matter using the fact that two dark matters participate in
  the pair-annihilation process (instead of one for the decay
  process).}
\begin{align}
  \Gamma_{\rm annihilation} =  n_X\langle\sigma v\rangle,
\end{align}
where $n_X$ is the number density of dark matter.  In addition, $\langle\sigma
v\rangle$ is the annihilation cross section, with which the Boltzmann
equation for the evolution of the number density of dark matter is
given by
\begin{align}
  \frac{d n_X}{d t} + 3 H n_X = 
  \langle\sigma v\rangle \left( n_{X,{\rm eq}}^2 - n_X^2 \right),
\end{align}
where $n_{X,{\rm eq}}$ is the equilibrium value of the number density
of dark matter.  We assume that the dark-matter annihilation occurs
through an $s$-wave process so that $\langle\sigma v\rangle$ is
independent of the relative velocity of dark-matter particles in the
non-relativistic limit.  The effects of the pair annihilation do not
change $n_X$ significantly during and after the BBN epoch because
those epochs are long after the freeze-out time of dark matter. Then
we use the following time-evolution of $n_X$:
\begin{align}
  n_X (t) = 
  \frac{3M_{\rm Pl}^2 H_0^2 \Omega_X}{m_X}
  \left( \frac{a(t)}{a_0} \right)^3,
\end{align}
where $M_{\rm Pl}\simeq 2.4\times 10^{18}~{\rm GeV}$ is the reduced
Planck scale, $H_0$ is the Hubble constant, $\Omega_X$ and $m_X$ are
the density parameter and the mass of dark matter, respectively, and
$a(t)$ and $a_0$ are the scale factor at the cosmic time $t$ and at
present, respectively.  (In our numerical calculation, we use
$H_0=68~{\rm km/sec/Mpc}$ and $\Omega_Xh^2=0.12$.)  $[dn_A/dt]_{\rm
  photodis}$, $[dn_A/dt]_{\rm hadrodis}$, and
$[dn_A/dt]_{p\leftrightarrow n}$ are proportional to
$n_X^2\langle\sigma v\rangle$, and the effects of the dark-matter
annihilation become more efficient as the annihilation cross section
$\langle\sigma v\rangle$ increases.  In order not to affect the
light-element abundances too much, $\langle\sigma v\rangle$ is bounded
from above.

Next, we summarize the new points in the calculation of the light
element abundances compared to Refs.~\cite{KKM,Kawasaki:2008qe}:
revision of the SBBN reaction rates and the treatment of the
hadrodissociations.  In the present study, we renewed reaction rates
by adopting the results of Ref.~\cite{Serpico:2004gx}.\footnote
{ For another recent study of nuclear reaction rates, see also
  Ref.~\cite{Xu:2013fha}, which we do not use in our analysis because
  there is a technical difficulty in implementing the errors given in
  Ref.~\cite{Xu:2013fha} into our Monte-Carlo analysis.  }
In order to take into account the uncertainties in the reaction rates,
we use the Monte-Carlo simulation to estimate the errors of the light
element abundances by assuming that the errors of the reaction rates
obey Gaussian distributions.

As for the
hadrodissociations, we have improved the treatment of the hadronic
showers initiated by injections of energetic hadrons  
into the thermal plasma,
including the following effects:\footnote
{For more details, see Ref.~\cite{KawKohMorTak}.}
\begin{itemize}
\item[1.] Effects of anti-nucleons emitted from dark-matter
  annihilation.
\item[2.] Effects of interconversion reactions between (anti-) neutron
  and (anti-) proton, with which injected and secondary-produced beam
  nucleons, as well as target nucleons, change their charges at the
  time of the inelastic scattering.
\end{itemize}
Concerning the effects of anti-nucleons, we have considered
scatterings of anti-nucleons off the background protons and $^4$He's.
The scatterings of anti-nucleons off the background protons produce
high-energy nucleons, which can destroy $^4$He's and produce copious
high-energy daughter particles (i.e., nucleons and light elements).
Although the inelastic scatterings of anti-nucleons off $^4$He's
produce high-energy daughter particles as well, we have neglected all
the final-state particles of the scatterings in the subsequent
calculation because of the lack of sufficient experimental data for
those inelastic scatterings.  It is expected that the inclusion of
such final-state particles would only make the BBN constraints
severer, and hence our treatment gives conservative bounds.
 
With the inclusion of the effects of anti-nucleons, the constraints
become stronger by 10$\%$ -- 30$\%$ for $m_X \sim $10 GeV -- 1 TeV. On
the other hand, by the latter effects (i.e., the effects of the
interconversions at inelastic scatterings), the constraints become
weaker by 50$\%$ -- 80$\%$. This is due to the fact that a high-energy
proton, which is interconverted from a beam neutron, tends to be
stopped more easily through electromagnetic interactions in the plasma
than the neutron with the same energy.

Now, we are at the position to discuss the bounds on the annihilation
cross section of dark matter $\langle\sigma v\rangle$.  The effects of
the dark-matter annihilation on the BBN depend on the final-state
particles produced by the annihilation process.  Here, we derive the
upper bound on $\langle\sigma v\rangle$, assuming that the
annihilation process is dominated by one of the following modes:
\begin{itemize}
\item $XX\rightarrow W^+W^-$,
\item $XX\rightarrow \bar{q}q$ (with $q=u,d,s,c,b$).
\end{itemize}
We consider only the cases where the pair annihilation results in the
production of two particles with an identical mass.  Thus, the energy
of individual final-state particles is equal to $m_X$.  For the
calculation of the photodissociation rate, it is necessary to acquire
the total amount of the energy injection in the form of electromagnetic
particles due to the pair annihilation.  In addition, the study of the
hadrodissociation and the $p\leftrightarrow n$ conversion
processes requires energy distributions of hadrons (in particular,
proton, neutron, and pions) produced by the annihilation of dark
matter.  The decay, cascade, and hadronization processes of the
standard-model particles are studied by using PYTHIA 8.2 package
\cite{Sjostrand:2014zea}.

\begin{figure}[t]
\vspace{-0. cm}
  \begin{center}
    \includegraphics[width=100mm]{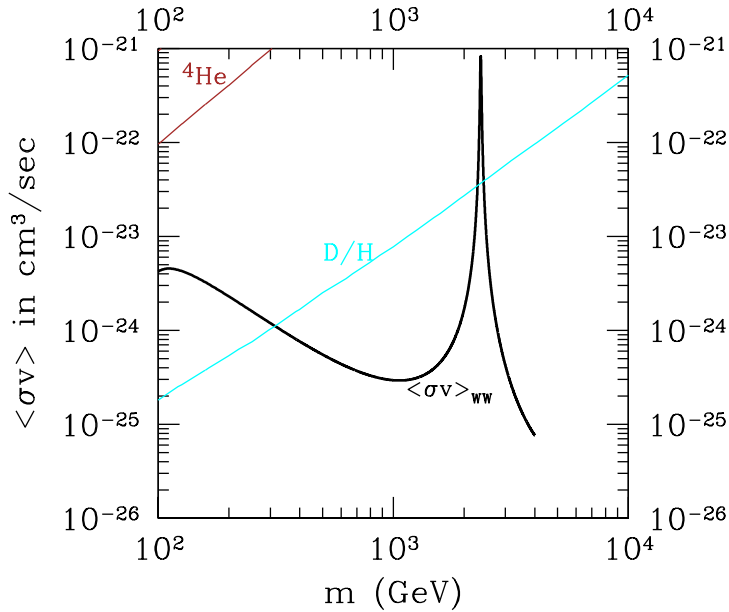}
  \end{center}
\vspace{-1. cm}
\caption{Upper bound at 95$\%$ C.L. on the annihilation cross section
  into the $W^+ W^-$ mode as a function of the dark-matter mass. The
  upper bound comes from the observational deuterium abundance. We
  also plot the theoretical prediction of the annihilation cross
  section for a pair of Winos into the $W^+ W^-$ mode taken from
  Ref.~\cite{Hisano:2008ti}.}
\label{fig:ww7.0}
\end{figure}

\begin{figure}[t]
\vspace{-0. cm}
  \begin{center}
    \includegraphics[width=100mm]{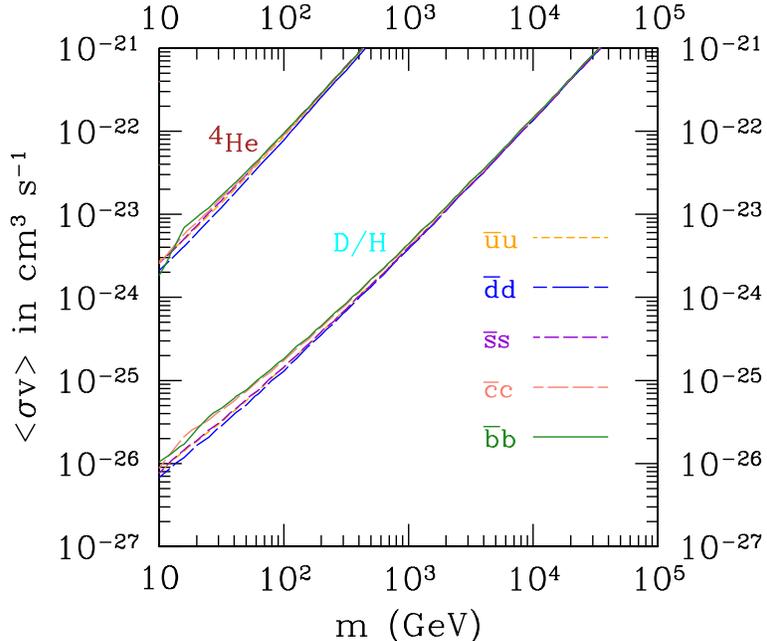}
  \end{center}
\vspace{-1. cm}
\caption{Upper bounds at 95$\%$ C.L. on the annihilation cross section
  into the $\bar{q} q$ mode ($\bar{u} u$, $\bar{d} d$, $\bar{s} s$,
  $\bar{c} c$, and $\bar{b} b$) as a function of the dark-matter
  mass. The upper bounds come from the observational deuterium
  abundance. }
\label{fig:mixed7.0}
\end{figure}

Comparing the theoretical predictions of light-element abundances with
observational constraints, we derive the bounds on the annihilation cross
section $\langle\sigma v\rangle$.  The results are shown in
Figs.~\ref{fig:ww7.0} and \ref{fig:mixed7.0} for $W^+ W^-$ and
$\bar{q}q$ emissions, respectively.\footnote
{We have also performed an analysis for the case 
where the dark matter dominantly annihilates into $ZZ$.  We have found that 
the bound is almost the same as the case with the 
$W^+ W^-$ final state.}
On the figures, only the constraints from D and $^4{\rm He}$ are shown
because that from $^3{\rm He/D}$ is too weak to show up in the
figures.  We can see that the constraints from D are much more
stringent than those from $^4{\rm He}$.

The upper bounds on $\langle\sigma v\rangle$ can be converted to
constraints on model parameters once the particle-physics model of
dark matter is specified.  First, let us consider the case where the
neutral Wino $\tilde{W}^0$ is dark matter in a supersymmetric model.
The neutral Wino is the superpartner of the neutral $SU(2)_L$ gauge
boson and one of the well-motivated candidates of dark matter, which
dominantly pair-annihilates into a $W^+W^-$ pair. In order for the
thermal relic Wino to be dark matter, the Wino mass is required to be
$\sim 3~{\rm TeV}$~\cite{Hisano:2006nn}.  However, Wino with smaller
mass can also account for dark matter if it is non-thermally
produced~\cite{Giudice:1998xp, Moroi:1999zb}.  In such a case, a
larger value of $\langle\sigma v\rangle$ is possible.  In Fig.\
\ref{fig:ww7.0}, we also plot the cross section for the process
$\tilde{W}^0\tilde{W}^0\rightarrow W^+W^-$ as a function of the
dark-matter mass (i.e., the Wino mass).  We can see that there exist
mass ranges in which the cross section is larger than the upper bound
obtained from the BBN. Assuming that neutral Wino is dark matter, the
Wino mass is constrained to be
\begin{align}
  320~{\rm GeV} \lesssim M_{\tilde{W}} \lesssim 2.3~{\rm TeV} ~~ {\rm or} ~~
  M_{\tilde{W}} \gtrsim 2.5~{\rm TeV}.
\end{align}
With the present analysis, the BBN bound on the Wino mass has become
more stringent than that obtained in the previous BBN analysis.
Neglecting the constraints from $^6$Li and $^7$Li, the bound was given
by $250~{\rm GeV} \lesssim M_{\tilde{W}} \lesssim 2.3~{\rm TeV}$ or
$M_{\tilde{W}} \gtrsim 2.4~{\rm TeV}$~\cite{Hisano:2008ti}.  We can
see that, in the Wino dark-matter scenario, the lowest possible value
of $M_{\tilde{W}}$ is significantly increased.  In addition, the
scenario of the thermal Wino dark matter (with $M_{\tilde{W}}\simeq 3\
{\rm TeV}$) is allowed.  Next, let us discuss the case of thermal
relic dark matter, assuming that the dark matter dominantly
annihilates into quark pairs.  In such a case, $\langle \sigma v
\rangle \sim 3 \times 10^{-26}~{\rm cm^3/sec}$ is required, which
results in $m_X \gtrsim 25 - 35~{\rm GeV}$.  There are some remarks
related with the recent reports of the possible $\gamma$-ray line
excess from the Galactic center at around GeV energies. In
Ref.~\cite{Daylan:2014rsa}, for example, it is claimed that the signal
can be fitted by the $\bar{q}q$ emission from annihilation of dark
matter with the mass of $36-51$~GeV.\footnote
{For other efforts, see also Refs.\ \cite{Hooperon}.}
Such a region is still consistent with the BBN bounds.

Finally, let us comment on other collider and astrophysical
constraints.  One constraint is from the direct search of charged Wino
at the LHC experiment.  Because the charged and neutral Winos are
almost mass-degenerate, the charged Wino becomes relatively long-lived
(i.e., $\tau_{\tilde{W}^\pm}\sim O(0.1~{\rm ns})$) if the neutral Wino
is the LSP~\cite{Feng:1999fu, Ibe:2012sx}. Then, the charged Wino
production may leave disappearing track at the LHC. The bound from the
disappearing-track search is currently $M_{\tilde{W}} \gtrsim 270~{\rm
  GeV}$ at 95\% C.L.~\cite{Aad:2013yna,CMS:2014gxa}, which is weaker
than the BBN bound.

Another constraint is estimated from the study of the energetic
$\gamma$-ray fluxes from Milky Way satellites.  In some case, sizable
$\gamma$-ray fluxes are expected from those satellites, although no
signal has been observed yet.  The negative search for the
$\gamma$-ray signal results in a lower bound on the mass of dark
matter.  For example, Ref.\ \cite{Bhattacherjee:2014dya} obtained the
constraint of $320\ {\rm GeV}\leq M_{\tilde{W}} \leq 2.25\ {\rm TeV}$
or $M_{\tilde{W}} \geq 2.43\ {\rm TeV}$ for the Wino dark-matter
scenario with the study of four nearby dwarf spheroidal galaxies
(dSphs) whose dark-matter profiles can be obtained with stellar
kinematic data.
In addition, with extra assumptions to determine the dark-matter profiles,
the Fermi-LAT collaboration \cite{Ackermann:2015zua} obtained $1~{\rm
  TeV} \lesssim M_{\tilde{W}} \lesssim 1.7~{\rm TeV}$ or
$M_{\tilde{W}} \gtrsim 2.8~{\rm TeV}$.  The Fermi-LAT collaboration
also obtained $m_X \gtrsim 110$ and $150~{\rm GeV}$ for the thermal
relic dark matter scenario with $W^+W^-$ and $\bar{q}q$ emissions,
respectively.  Although these constraints are comparable to or severer
than the BBN ones, constraints from $\gamma$-ray in general may suffer
from uncertainties of the density profiles of dark matter in galaxies.

The recent observation of the anti-proton flux in high-energy cosmic
ray~\cite{AMS02} also puts a bound on the mass of dark
matter~\cite{AMS02DMLimits}.  Importantly, however, the theoretical
prediction of the anti-proton flux from the dark-matter annihilation
has large uncertainty because the anti-proton flux is sensitive to the
model of cosmic-ray propagation as well as the dark-matter density
profile of the Milky Way.  Using the result of
Ref.~\cite{Evoli:2015vaa}, for example, the anti-proton constraint on
the Wino dark matter scenario is $M_{\tilde{W}} \gtrsim 240-250~{\rm
  GeV}$.  In addition, for the case of $\bar{q}q$ emission for the
annihilation process, the mass of dark matter is constrained to be
$m_X \gtrsim 20~{\rm GeV}$, assuming the annihilation cross section to
realize the thermal relic dark matter scenario.  These constraints are
weaker than the BBN constraints obtained by our analysis. (However,
they may become more stringent if one takes propagation models or
density profiles other than that adopted in
Ref.~\cite{Evoli:2015vaa}.)

Since dark-matter annihilation affects the ionization history of the
universe, the observations of the cosmic microwave background (CMB)
constrain the annihilation cross section.  The recent Planck obtained
$ \langle \sigma v\rangle \lesssim 4\times 10^{-28}\text{
  cm}^3\text{s}^{-1}f^{-1}(m_X/\text{GeV})$~\cite{Ade:2015xua}, where
$f$ is the fraction of the injected energy that goes into the
background plasma.  $f$ is expected to be $\sim O(0.1)$; for example
$f\simeq 0.15$ for the annihilation into $\tau^{+}\tau^{-}$ and
$f\simeq 0.2$ for the annihilation into $\mu^{+}\mu^{-}$ or
$b\bar{b}$~\cite{Ade:2015xua}.  If we take $f\sim 0.2$ as a typical
value, the Planck result leads to the constraints, $400~\text{GeV}
\lesssim M_{\tilde{W}} \lesssim 2.1~\text{TeV}$ or $M_{\tilde{W}}
\gtrsim 2.6~\text{TeV}$ for the Wino dark matter, and $m_{X} \gtrsim
15~\text{GeV}$ for the thermal relic dark matter with $q\bar{q}$
emission.  (For more accurate bounds, precise calculation of $f$
is necessary for each model.)  Therefore, the CMB constraints are
comparable to the BBN constraints.

In summary, we have studied effects of dark-matter annihilation in the
early universe on the abundances of the light elements synthesized
during the BBN epoch.  If the dark-matter annihilation results in the
production of electromagnetic and hadronic particles, they affect the
abundances of the light elements through photodissociation,
hadrodissociation, and $p\leftrightarrow n$ conversion processes.  We
have calculated the abundances of D, ${\rm ^3He}$, and ${\rm ^4He}$,
taking into account above processes.  In the study of the effects of
dark-matter annihilation, we have improved the treatment of the
hadrodissociation processes. In particular, (i) we have included the
effects of the anti-nucleon emitted by the annihilation process of dark
matter, and (ii) we take account of the interconversion reactions
between neutron and proton at inelastic scatterings.  Then,
comparing the theoretical prediction with the latest observational
constraints on the primordial abundances of the light elements, we have
derived the upper bounds on the pair annihilation cross section of
dark matter.  We found that the latest results on the measurements of
the D abundance, which result in a precise determination of the
primordial abundance of D, have a strong impact.  For the case of Wino
dark matter, for example, the BBN constraint requires the Wino mass to
be $320~{\rm GeV} \lesssim M_{\tilde{W}} \lesssim 2.3~{\rm TeV}$ or 
$M_{\tilde{W}} \gtrsim 2.5~{\rm TeV}$.

\vspace{1cm} We thank M. Arnould and P.~D. Serpico for useful
discussions. This work is supported in part by Grant-in-Aid for
Scientific research from the Ministry of Education, Science, Sports,
and Culture (MEXT), Japan, Nos.\ 23104008 (T.M. and Y.T.), 25400248
(M.K.), 26105520 (K.K.), 26247042 (K.K.), 26400239 (T.M.) and
15H05889 (M.K. and K.K.). The work of K.K. is also supported by the
Center for the Promotion of Integrated Science (CPIS) of Sokendai
(1HB5804100) and World Premier International Research Center
Initiative (WPI Initiative), MEXT, Japan (M.K. and T.M.).



\begin{thebibliography}{99}

\bibitem{Ade:2015xua} 
  P.~A.~R.~Ade {\it et al.} [Planck Collaboration],
  arXiv:1502.01589 [astro-ph.CO].

\bibitem{Reno:1987qw}
  M.~H.~Reno and D.~Seckel,
  Phys.\ Rev.\ D {\bf 37}, 3441 (1988).

\bibitem{Frieman:1989fx} 
  J.~A.~Frieman, E.~W.~Kolb and M.~S.~Turner,
  Phys.\ Rev.\ D {\bf 41}, 3080 (1990),

\bibitem{Jedamzik:2004ip} 
  K.~Jedamzik,
  Phys.\ Rev.\ D {\bf 70}, 083510 (2004)
  [astro-ph/0405583].

\bibitem{Hisano:2008ti} 
  J.~Hisano, M.~Kawasaki, K.~Kohri and K.~Nakayama,
  Phys.\ Rev.\ D {\bf 79}, 063514 (2009)
  [arXiv:0810.1892 [hep-ph]];
  Phys.\ Rev.\ D {\bf 80}, 029907 (2009)
  [arXiv:0810.1892 [hep-ph]].

\bibitem{Hisano:2009rc} 
  J.~Hisano, M.~Kawasaki, K.~Kohri, T.~Moroi and K.~Nakayama,
  Phys.\ Rev.\ D {\bf 79}, 083522 (2009)
  [arXiv:0901.3582 [hep-ph]].

\bibitem{Hisano:2011dc} 
  J.~Hisano, M.~Kawasaki, K.~Kohri, T.~Moroi, K.~Nakayama and T.~Sekiguchi,
  Phys.\ Rev.\ D {\bf 83}, 123511 (2011).
  [arXiv:1102.4658 [hep-ph]].

\bibitem{Henning:2012rm}
  B.~Henning and H.~Murayama,
  arXiv:1205.6479 [hep-ph].

\bibitem{Nollett:2013pwa} 
  K.~M.~Nollett and G.~Steigman,
  Phys.\ Rev.\ D {\bf 89}, no. 8, 083508 (2014)
  [arXiv:1312.5725 [astro-ph.CO]];
  Phys.\ Rev.\ D {\bf 91}, no. 8, 083505 (2015)
  [arXiv:1411.6005 [astro-ph.CO]].

\bibitem{Cooke:2013cba} 
  R.~Cooke, M.~Pettini, R.~A.~Jorgenson, M.~T.~Murphy and C.~C.~Steidel,
  arXiv:1308.3240 [astro-ph.CO].

\bibitem{Izotov:2014fga} 
  Y.~I.~Izotov, T.~X.~Thuan and N.~G.~Guseva,
  Mon.\ Not.\ Roy.\ Astron.\ Soc.\  {\bf 445}, 778 (2014)
  [arXiv:1408.6953 [astro-ph.CO]].

\bibitem{Aver:2015iza} 
  E.~Aver, K.~A.~Olive and E.~D.~Skillman,
  JCAP {\bf 1507}, no. 07, 011 (2015)
  [arXiv:1503.08146 [astro-ph.CO]].

\bibitem{GG03}
  J. Geiss and  G. Gloeckler, 
  Space Sience Reviews {\bf 106}, 3 (2003).

\bibitem{Sbordone:2010zi} 
  L.~Sbordone, P.~Bonifacio, E.~Caffau, H.-G.~Ludwig, N.~T.~Behara, J.~I.~G.~Hernandez, M.~Steffen and R.~Cayrel {\it et al.},
  Astron.\ Astrophys.\  {\bf 522}, A26 (2010)
  [arXiv:1003.4510 [astro-ph.GA]].

\bibitem{KKM} 
  M.~Kawasaki, K.~Kohri and T.~Moroi,
  Phys.\ Lett.\ B {\bf 625}, 7 (2005)
  [astro-ph/0402490];
  Phys.\ Rev.\ D {\bf 71}, 083502 (2005)
  [astro-ph/0408426].

\bibitem{Kawasaki:2008qe}
  M.~Kawasaki, K.~Kohri, T.~Moroi and A.~Yotsuyanagi,
  Phys.\ Rev.\ D {\bf 78} (2008) 065011
  [arXiv:0804.3745 [hep-ph]].

\bibitem{Serpico:2004gx} 
  P.~D.~Serpico, S.~Esposito, F.~Iocco, G.~Mangano, G.~Miele and O.~Pisanti,
  JCAP {\bf 0412}, 010 (2004)
  [astro-ph/0408076].

\bibitem{Xu:2013fha} 
  Y.~Xu, K.~Takahashi, S.~Goriely, M.~Arnould, M.~Ohta and H.~Utsunomiya,
  Nucl.\ Phys.\ A {\bf 918}, 61 (2013)
  [arXiv:1310.7099 [nucl-th]].

\bibitem{KawKohMorTak}
  M.~Kawasaki, K.~Kohri, T.~Moroi, and Y.~Takaesu,
  work in progress.

\bibitem{Sjostrand:2014zea} 
  T.~Sj\"ostrand {\it et al.},
  Comput.\ Phys.\ Commun.\  {\bf 191}, 159 (2015)
  [arXiv:1410.3012 [hep-ph]].

\bibitem{Hisano:2006nn}
  J.~Hisano, S.~Matsumoto, M.~Nagai, O.~Saito and M.~Senami,
  Phys.\ Lett.\ B {\bf 646} (2007) 34
  [hep-ph/0610249].

\bibitem{Giudice:1998xp}
  G.~F.~Giudice, M.~A.~Luty, H.~Murayama and R.~Rattazzi,
  JHEP {\bf 9812} (1998) 027
  [hep-ph/9810442].

\bibitem{Moroi:1999zb}
  T.~Moroi and L.~Randall,
  Nucl.\ Phys.\ B {\bf 570} (2000) 455
  [hep-ph/9906527].

\bibitem{Daylan:2014rsa} 
  T.~Daylan, D.~P.~Finkbeiner, D.~Hooper, T.~Linden, S.~K.~N.~Portillo, N.~L.~Rodd and T.~R.~Slatyer,
  arXiv:1402.6703 [astro-ph.HE].

\bibitem{Hooperon}
L.~Goodenough and D.~Hooper,
  arXiv:0910.2998 [hep-ph];
D.~Hooper and L.~Goodenough,
  Phys.\ Lett.\ B {\bf 697}, 412 (2011)
  [arXiv:1010.2752 [hep-ph]];
D.~Hooper and T.~Linden,
  Phys.\ Rev.\ D {\bf 84}, 123005 (2011)
  [arXiv:1110.0006 [astro-ph.HE]];
K.~N.~Abazajian and M.~Kaplinghat,
  Phys.\ Rev.\ D {\bf 86}, 083511 (2012)
  [Phys.\ Rev.\ D {\bf 87}, 129902 (2013)]
  [arXiv:1207.6047 [astro-ph.HE]];
D.~Hooper and T.~R.~Slatyer,
  Phys.\ Dark Univ.\  {\bf 2}, 118 (2013)
  [arXiv:1302.6589 [astro-ph.HE]];
C.~Gordon and O.~Macias,
  Phys.\ Rev.\ D {\bf 88}, no. 8, 083521 (2013)
  [Phys.\ Rev.\ D {\bf 89}, no. 4, 049901 (2014)]
  [arXiv:1306.5725 [astro-ph.HE]];
W.~C.~Huang, A.~Urbano and W.~Xue,
  arXiv:1307.6862 [hep-ph];
 K.~N.~Abazajian, N.~Canac, S.~Horiuchi and M.~Kaplinghat,
  Phys.\ Rev.\ D {\bf 90}, no. 2, 023526 (2014)
  [arXiv:1402.4090 [astro-ph.HE]];
  T.~Daylan, D.~P.~Finkbeiner, D.~Hooper, T.~Linden, S.~K.~N.~Portillo, N.~L.~Rodd and T.~R.~Slatyer,
  arXiv:1402.6703 [astro-ph.HE].

\bibitem{Feng:1999fu}
  J.~L.~Feng, T.~Moroi, L.~Randall, M.~Strassler and S.~f.~Su,
  Phys.\ Rev.\ Lett.\  {\bf 83} (1999) 1731
  [hep-ph/9904250].

\bibitem{Ibe:2012sx}
  M.~Ibe, S.~Matsumoto and R.~Sato,
  Phys.\ Lett.\ B {\bf 721} (2013) 252
  [arXiv:1212.5989 [hep-ph]].

\bibitem{Aad:2013yna} 
  G.~Aad {\it et al.} [ATLAS Collaboration],
  Phys.\ Rev.\ D {\bf 88}, no. 11, 112006 (2013)
  [arXiv:1310.3675 [hep-ex]].

\bibitem{CMS:2014gxa} 
  V.~Khachatryan {\it et al.} [CMS Collaboration],
  JHEP {\bf 1501}, 096 (2015)
  [arXiv:1411.6006 [hep-ex]].

\bibitem{Bhattacherjee:2014dya}
  B.~Bhattacherjee, M.~Ibe, K.~Ichikawa, S.~Matsumoto and K.~Nishiyama,
  JHEP {\bf 1407} (2014) 080
  [arXiv:1405.4914 [hep-ph]].

\bibitem{Ackermann:2015zua} 
  M.~Ackermann {\it et al.} [Fermi-LAT Collaboration],
  arXiv:1503.02641 [astro-ph.HE].

\bibitem{AMS02}
  AMS-02 collaboration, 
  talks at the ``AMS DAYS AT CERN -- The Future of Cosmic Ray Physics
  and Latest Results,'' April 15-17, 2015, CERN.

\bibitem{AMS02DMLimits}
  G.~Giesen, M.~Boudaud, Y.~Genolini, V.~Poulin, M.~Cirelli, P.~Salati and P.~D.~Serpico,
  arXiv:1504.04276 [astro-ph.HE];
  H.~B.~Jin, Y.~L.~Wu and Y.~F.~Zhou,
  arXiv:1504.04604 [hep-ph];
  C.~Evoli, D.~Gaggero and D.~Grasso,
  arXiv:1504.05175 [astro-ph.HE];
  M.~Ibe, S.~Matsumoto, S.~Shirai and T.~T.~Yanagida,
  Phys.\ Rev.\ D {\bf 91}, no. 11, 111701 (2015)
  [arXiv:1504.05554 [hep-ph]];
  K.~Hamaguchi, T.~Moroi and K.~Nakayama,
  Phys.\ Lett.\ B {\bf 747}, 523 (2015)
  [arXiv:1504.05937 [hep-ph]];
  S.~J.~Lin, X.~J.~Bi, P.~F.~Yin and Z.~H.~Yu,
  arXiv:1504.07230 [hep-ph].

\bibitem{Evoli:2015vaa}
  C.~Evoli, D.~Gaggero and D.~Grasso,
  arXiv:1504.05175 [astro-ph.HE].

\end{thebibliography}
\end{document}